\documentclass[12pt]{article}
\usepackage{graphicx}
\usepackage{citehack,cite}

\newcommand{\non}{\nonumber \\}
\newcommand{\ve}[1]{{\bf #1}}
\newcommand{\be}{\begin{equation}}
\newcommand{\ee}{\end{equation}}
\newcommand{\bea}{\begin{eqnarray}}
\newcommand{\eea}{\end{eqnarray}}
\newcommand{\sli}{\sum\limits}
\newcommand{\ili}{\int\limits}
\newcommand{\lp}{\left (}
\newcommand{\rp}{\right )}
\newcommand{\lb}{\left \{}

\newcommand{\lbr}{\left [}
\newcommand{\rbr}{\right ]}
\newcommand{\ld}{\left .}
\newcommand{\rd}{\right .}
\newcommand{\rhok}{\rho_{\ve{k}}}
\newcommand{\rhomk}{\rho_{-\ve{k}}}
\newcommand{\omk}{\omega_{\ve{k}}}
\newcommand{\cM}{{\cal M}}
\newcommand{\cW}{{\cal W}}
\newcommand{\mt}{m_{\tau}}
\newcommand{\np}{n_p}
\newcommand{\npm}{n_p+m}
\newcommand{\npmb}{n_p+m+1}
\newcommand{\nps}{n_p'}
\newcommand{\npsb}{n_p'-1}
\newcommand{\Delb}{\Delta_1}
\newcommand{\hhcqp}{\tilde h^2+\tilde h_c^2}
\newcommand{\hhcqf}{\frac{\tilde h^2}{\tilde h_c^2}}
\newcommand{\hchqf}{\frac{\tilde h_c^2}{\tilde h^2}}
\newcommand{\ccaaHd}{c_{20}^{(0)}H_3}
\newcommand{\faHc}{f_0H_c}
\newcommand{\btPhia}{\beta\tilde{\Phi}(0)}
\newcommand{\btPhik}{\beta\tilde{\Phi}(k)}
\newcommand{\tPhik}{\tilde{\Phi}(k)}
\newcommand{\tPhia}{\tilde{\Phi}(0)}
\newcommand{\half}{\frac{1}{2}}

\begin{document}

\begin{center}
{\bf METHOD OF CALCULATING THE FREE ENERGY OF THREE-DIMENSIONAL
ISING-LIKE SYSTEM IN AN EXTERNAL FIELD WITH THE USE OF THE $\rho^6$
MODEL}
\end{center}

\begin{center}
{\sc I.V. Pylyuk, M.P. Kozlovskii}
\end{center}

\begin{center}
{\it Institute for Condensed Matter Physics  \\
of the National Academy of Sciences of Ukraine, \\
1~Svientsitskii Str., UA-79011 Lviv, Ukraine} \\
E-mail: piv@icmp.lviv.ua
\end{center}

\vspace{0.5cm}

{\small
The microscopic approach to calculating the free energy of a
three-dimensional Ising-like system in a homogeneous external field is
developed in the higher non-Gaussian approximation (the $\rho^6$ model)
at temperatures above the critical value of $T_c$ ($T_c$ is the
phase-transition temperature in the absence of an external field).
The free energy of the system is found by separating the contributions
from the short- and long-wave spin-density oscillation modes taking into
account both the temperature and field fluctuations of the order
parameter. Our analytical calculations do not involve power series in
the scaling variable and are valid in the whole field-temperature plane
near the critical point including the region in the vicinity of the
limiting field $\tilde h_c$, which divides external fields into the weak
and strong ones (i.e., the crossover region). In this region, the
temperature and field effects on the system are equivalent, the scaling
variable is of the order of unity, and power series are not efficient.
The obtained expression for the free energy contains the leading terms
and terms determining the temperature and field confluent corrections.
}

\vspace{0.5cm}


\section{Introduction}

Despite the great successes in the investigation of three-dimensional
($3D$) Ising-like systems made by means of various methods (see, for
example, \cite{pv102}), the statistical description of the critical
behavior of the mentioned systems in terms of the temperature and field
variables and the calculation of scaling functions are still of
interest \cite{efs103}. The Ising model is widely used in the theory of
phase transitions for the study of the properties of various magnetic
and non-magnetic systems (ferroelectrics, ferromagnets, binary mixtures,
etc.).

The description of the phase transitions in the $3D$ magnets, usually, is
associated with the absence of exact solutions and with many approximate
approaches for obtaining different system characteristics.
In this article, the behavior of a $3D$ Ising-like system near the
critical point in a homogeneous external field is studied using the
collective variables (CV) method \cite{ymo287,rev9789,ykpmo101}. The main
peculiarity of this method is the integration of short-wave spin-density
oscillation modes, which is generally done without using perturbation
theory. The CV method is similar to the Wilson non-perturbative
renormalization-group (RG) approach (integration on fast modes and
construction of an effective theory for slow modes)
\cite{btw102,tw194,bb101}. The term collective variables is a common name
for a special class of variables that are specific for each individual
physical system \cite{ymo287,rev9789}. The CV set contains variables
associated with order parameters. Because of this, the phase space of CV
is most natural for describing a phase transition. For magnetic systems,
the CV $\rhok$ are the variables associated with modes of spin-moment
density oscillations, while the order parameter is related to the
variable $\rho_0$, in which the subscript ``0'' corresponds to the peak
of the Fourier transform of the interaction potential.

The free energy of a $3D$ Ising-like system in an external field at
temperatures above $T_c$ is calculated using the non-Gaussian
spin-density fluctuations, namely the sextic measure density. The latter
is represented as an exponential function of the CV whose argument
includes the powers with the corresponding coupling constants up to the
sixth power of the variable (the $\rho^6$ model).

The present publication supplements the earlier
works \cite{kpd297,p599,ykp202,ypk102}, in which
the $\rho^6$ model was used for calculating the free energy and other
thermodynamic functions of the system in the absence of an external
field. The $\rho^6$ model provides a better quantitative description of
the critical behavior of a $3D$ Ising-like magnet than the $\rho^4$
model \cite{ykp202}. For each of the $\rho^{2m}$ models, there exists
a preferred value of the RG parameter $s = s^*$ ($s^* = 3.5862$ for the
$\rho^4$ model, $s^* = 2.7349$ for the $\rho^6$ model, $s^* = 2.6511$ for
the $\rho^8$ model, and $s^* = 2.6108$ for the $\rho^{10}$ model)
nullifying the average value of the coefficient in the term with the second
power in the effective density of measure at the fixed point. The values of
$s$ close to $s^*$ are optimal for the given method
of calculations. The difference form of the recurrence relations (RR)
between the coefficients of effective non-Gaussian densities of measures
operates successfully just in this region of $s$.
It was established (see, for example, \cite{ykp202,ypk102}) that as the
form of the density of measure becomes more complicated, the dependence of
the critical exponent of the correlation length $\nu$
on the RG parameter $s$ becomes weaker gradually, and, starting from
the sextic density of measure, the value of the exponent $\nu$, having a
tendency to saturate with increasing $m$ (which characterizes the order
of the $\rho^{2m}$ model, $m = 2,3,4,5$), changes insignificantly.
The Ising model corresponds to the $\rho^{2m}$ model approximation,
where the order of the model $2m \geq 4$.
The $\rho^4$ model allows us to go beyond the classical
analysis and to describe all qualitative aspects of the second-order
phase transition. The critical behavior of a $3D$ Ising-like system
within the CV method can be described quantitatively at $2m \geq 6$, and,
in particular, at $2m = 6$. It was shown in \cite{ykp202,ypk202} that
the graphs of the temperature dependences of the order parameter
(the spontaneous magnetization) and specific heat for the $\rho^6$ model
agree more closely with the Liu and Fisher's results \cite{lf189} than
the corresponding plots for the $\rho^4$ model.
The correctness of the choice of
the $\rho^6$ model for investigations is also confirmed in
\cite{t294} and \cite{t397}, where the effective potential is studied for
the scalar field theory in three dimensions in
the symmetric and spontaneously broken phases, respectively. In this
case, probability distributions of average magnetization in the $3D$ Ising
model in an external field, obtained with the help of the Monte Carlo
method, were used. Tsypin \cite{t294,t397} proved that the term with
the sixth power of the variable in the effective potential plays
an important role.

The methods existing at present make it possible to calculate universal
quantities to a quite high degree of accuracy (see, for example,
\cite{pv102}). The advantage of the CV method lies in the possibility of
obtaining and analysing thermodynamic characteristics as functions of the
microscopic parameters of the initial system \cite{ykp202,ypk102,ypk202}.
The results of calculations for a $3D$ Ising system on the basis of the
$\rho^4$ and $\rho^6$ models are in accord with the results obtained by
other authors (see \cite{ykp202,ypk202}). In \cite{kpp606}, the scaling
functions of the order parameter and susceptibility, calculated on the
basis of the free energy for the $\rho^4$ model, were graphically
compared with other authors' data. Our results accord with the results
obtained within the framework of the parametric representation of the
equation of state \cite{gz197} and Monte Carlo simulations \cite{efs103}.

The expressions for the thermodynamic
characteristics of the system in the presence of an external field
have already been obtained on the basis of
the simplest non-Gaussian measure density (the $\rho^4$ model) in
\cite{pkp105,kpp406,kpp506,kpp305} using the point of exit of the system
from the critical regime as a function of the temperature (the weak-field
region) or of the field (the strong-field region). In
\cite{pkp105,kpp406}, the thermodynamic characteristics are presented in
the form of series expansions in the variables, which are combinations
of the temperature and field. Our calculations in the $\rho^4$ model
approximation were also performed for
temperatures $T>T_c$ \cite{kpp506} and $T<T_c$ \cite{kpp305} without
using similar expansions for the roots of cubic equations appearing in
the theoretical analysis. In this article, the free energy of a $3D$
uniaxial magnet within the framework of the more complicated $\rho^6$
model is found without using series expansions introducing the
generalized point of exit of the system from the critical regime. This
point takes into account the temperature and field variables
simultaneously. In our earlier article \cite{kpp606}, the point of exit
of the system from the critical regime was found in the simpler
non-Gaussian approximation (the $\rho^4$ model) using the numerical
calculations. In contrast to \cite{kpp606}, the point of exit of the
system in the present article is explicitely defined as a function of
the temperature and field. This allows one to obtain the free energy
without involving the numerical calculations that is our problem
solved in the present article.

\section{Integration of partition function of the system in the $\rho^6$
model approximation taking into account effect of an external field}

We consider a $3D$ Ising-like system on a simple cubic lattice
with $N$ sites and period $c$ in a homogeneous external field $h$.
The Hamiltonian of such a system has the form
\be
H=-\half~\sli_{\ve{j},\ve{l}}\Phi(r_{\ve{j}\ve{l}})\sigma_{\ve{j}}
\sigma_{\ve{l}}-h\sli_{\ve{j}}\sigma_{\ve{j}},
\label{ft1}
\ee
where $r_{\ve{j}\ve{l}}$ is the distance between particles at sites
$\ve{j}$ and $\ve{l}$, and $\sigma_{\ve{j}}$ is the operator of the $z$
component of spin at the $\ve{j}$th site, having two eigenvalues +1
and -1. The interaction potential is an exponentially decreasing function
\be
\Phi(r_{\ve{j}\ve{l}})=A\exp\lp-\frac{r_{\ve{j}\ve{l}}}{b}\rp.
\label{ft2}
\ee
Here $A$ is a constant and $b$ is the radius of effective interaction.
For the Fourier transform of the interaction potential, we use the
following approximation \cite{ymo287,ykp202,ypk102}:
\be
\tPhik=\lb
\begin{array}{ll}
\tPhia(1-2b^2k^2), & k\leq B', \\
0, & B'<k\leq B,
\end{array}
\rd
\label{ft3}
\ee
where $B$ is the boundary of the Brillouin half-zone ($B=\pi/c$),
$B'=(b\sqrt{2})^{-1}$, $\tPhia=8\pi A(b/c)^3$.

In the CV representation for the partition function of the system,
we have \cite{ymo287,dkp196}
\be
Z=\int\exp\lbr\half~\sli_{\ve{k}}\btPhik\rhok\rhomk+
\beta h\sqrt{N}\rho_0\rbr J(\rho)~(d\rho)^N.
\label{ft4}
\ee
Here the summation over the wave vectors $\ve{k}$ is carried out within
the first Brillouin zone, $\beta=1/(kT)$ is the inverse thermodynamic
temperature, the CV $\rhok$ are introduced by means of the functional
representation for operators of spin-density oscillation modes
$\hat\rhok=(\sqrt{N})^{-1}\sum_{\ve{l}}\sigma_{\ve{l}}\exp(-i\ve{k}\ve{l})$,
\bea
J(\rho)&=&2^N\int\exp\lbr 2\pi i~\sli_{\ve{k}}\omk\rhok+
~\sli_{n\geq 1}\lp 2\pi i\rp^{2n} N^{1-n}\rd\non
& & \ld\times\frac{\cM_{2n}}{(2n)!}~\sli_{\ve{k}_1,\ldots,\ve{k}_{2n}}
\omega_{\ve{k}_1}\cdots\omega_{\ve{k}_{2n}}~\delta_{\ve{k}_1+\cdots+
\ve{k}_{2n}}\rbr~(d\omega)^N\non
\label{ft5}
\eea
is the Jacobian of transition from the set of $N$ spin variables
$\sigma_{\ve{l}}$ to the set of CV $\rhok$, and
$\delta_{\ve{k}_1+\cdots+\ve{k}_{2n}}$ is the Kronecker symbol.
The variables $\omk$ are conjugate to $\rhok$, and the cumulants
$\cM_{2n}$ assume constant values (see \cite{ymo287,rev9789,ykpmo101}).

Proceeding from Eqs. (\ref{ft4}) and (\ref{ft5}), we obtain the
following initial expression for the partition function of the system in
the $\rho^6$ model approximation:
\bea
Z&=&2^N2^{(N'-1)/2}e^{a'_0N'}\int\exp\Biggl[ -a'_1(N')^{1/2}\rho_0\non
& & -\half~\sli_{{\ve{k}}\atop{k\leq B'}}d'(k)\rhok\rhomk
-~\sli_{l=2}^3\frac{a'_{2l}}{(2l)!(N')^{l-1}}\non
& & \times~\sli_{{\ve{k}_1,\ldots,\ve{k}_{2l}}\atop{k_i\leq B'}}
\rho_{\ve{k}_1}\cdots\rho_{\ve{k}_{2l}}~\delta_{\ve{k}_1+\cdots+
\ve{k}_{2l}}\Biggr]~(d\rho)^{N'}.
\label{ft6}
\eea
Here $N'=Ns_0^{-d}$ ($d=3$ is the space dimension),
$s_0=B/B'=\pi\sqrt{2} b/c$, and $a'_1=-s_0^{d/2}h'$, $h'=\beta h$. The
expressions for the remaining coefficients are given in
\cite{kpd297,p599,ykp202,ypk102}. These coefficients are functions of
$s_0$, i.e., of the ratio of microscopic parameters $b$ and $c$.
The integration over the zeroth, first, second, $\ldots$, $n$th layers
of the CV phase space \cite{ymo287,rev9789,ykpmo101,ykp202} leads to
the representation of the partition function in the form of a product of
the partial partition functions $Q_n$ of individual layers and the
integral of the ``smoothed'' effective measure density
\bea
Z&=&2^N2^{(N_{n+1}-1)/2}Q_0Q_1\cdots Q_n[Q(P_n)]^{N_{n+1}}\non
& & \times\int\cW_6^{(n+1)}(\rho)~(d\rho)^{N_{n+1}}.
\label{ft7}
\eea
The expressions for ~$Q_n$, $Q(P_n)$ ~are presented in
\cite{kpd297,p599,ykp202,ypk102}, and ~$N_{n+1}=N's^{-d(n+1)}$.
The sextic measure density of the ($n+1$)th block
structure $\cW_6^{(n+1)}(\rho)$ has the form
\bea
\lefteqn{\cW_6^{(n+1)}(\rho)}\non
& & =\exp\lbr-a_1^{(n+1)}N_{n+1}^{1/2}\rho_0
-\half~\sli_{{\ve{k}}\atop{k\leq B_{n+1}}}d_{n+1}(k)\rhok\rhomk\rd\non
& & \ \ \ld-\sli_{l=2}^3\frac{a_{2l}^{(n+1)}}{(2l)!N_{n+1}^{l-1}}
\sli_{{\ve{k}_1,\ldots,\ve{k}_{2l}}\atop{k_i\leq B_{n+1}}}
\rho_{\ve{k}_1}\cdots\rho_{\ve{k}_{2l}}~\delta_{\ve{k}_1+\cdots+
\ve{k}_{2l}}\rbr,
\label{ft8}
\eea
where $B_{n+1}=B's^{-(n+1)}$, $d_{n+1}(k)=a_2^{(n+1)}-\btPhik$,
$a_1^{(n+1)}$ and $a_{2l}^{(n+1)}$ are the renormalized values of the
coefficients $a'_1$ and $a'_{2l}$ after integration over $n+1$ layers of
the phase space of CV. The coefficients $a_1^{(n)}=s^{-n}t_n$,
$d_n(0)=s^{-2n}r_n$ [appearing in $d_n(k)=d_n(0)+2\btPhia b^2k^2$],
$a_4^{(n)}=s^{-4n}u_n$, and $a_6^{(n)}=s^{-6n}w_n$ are connected with the
coefficients of the ($n+1$)th layer through the RR
\bea
& & t_{n+1}=s^{(d+2)/2} t_n,\non
& & r_{n+1}=s^2\lbr -q+u_n^{1/2}Y(h_n,\alpha_n)\rbr,\non
& & u_{n+1}=s^{4-d} u_n B(h_n,\alpha_n),\non
& & w_{n+1}=s^{6-2d} u_n^{3/2}D(h_n,\alpha_n)
\label{ft9}
\eea
whose solutions
\bea
t_n&=&t^{(0)}-s_0^{d/2}h'E_1^n,\non
r_n&=&r^{(0)}+c_1E_2^n+c_2w_{12}^{(0)}(u^{(0)})^{-1/2}E_3^n\non
& & +c_3w_{13}^{(0)}(u^{(0)})^{-1}E_4^n,\non
u_n&=&u^{(0)}+c_1w_{21}^{(0)}(u^{(0)})^{1/2}E_2^n+c_2E_3^n\non
& & +c_3w_{23}^{(0)}(u^{(0)})^{-1/2}E_4^n,\non
w_n&=&w^{(0)}+c_1w_{31}^{(0)}u^{(0)}E_2^n\non
& & +c_2w_{32}^{(0)}(u^{(0)})^{1/2}E_3^n+c_3E_4^n
\label{ft10}
\eea
in the region of the critical regime are used for calculating the free
energy of the system. Here
\bea
& & Y(h_n,\alpha_n)=s^{d/2}F_2(\eta_n,\xi_n)
\lbr C(h_n,\alpha_n)\rbr^{-1/2},\non
& & B(h_n,\alpha_n)=s^{2d}C(\eta_n,\xi_n)
\lbr C(h_n,\alpha_n)\rbr^{-1},\non
& & D(h_n,\alpha_n)=s^{7d/2}N(\eta_n,\xi_n)
\lbr C(h_n,\alpha_n)\rbr^{-3/2}.
\label{ft11}
\eea
The quantity $q=\bar q\btPhia$ determines the average value of the
Fourier transform of the potential
$\beta\tilde{\Phi}(B_{n+1},B_n)=\btPhia-q/s^{2n}$ in
the $n$th layer (in this article, $\bar q=(1+s^{-2})/2$ corresponds to
the arithmetic mean value of $k^2$ on the interval $(1/s,1]$). The basic
arguments $h_n$ and $\alpha_n$ are determined by the coefficients of the
sextic measure density of the $n$th block structure. The intermediate
variables $\eta_n$ and $\xi_n$ are functions of $h_n$ and $\alpha_n$. The
expressions for both basic and intermediate arguments as well as the
special functions appearing in Eqs. (\ref{ft11}) are the same as in
the absence of an external field (see \cite{kpd297,p599,ykp202,ypk102}).
The quantities $E_l$ in Eqs. (\ref{ft10}) are the eigenvalues of
the matrix of the RG linear transformation
\be
\lp
\begin{array}{c}
t_{n+1}-t^{(0)} \\
r_{n+1}-r^{(0)} \\
u_{n+1}-u^{(0)} \\
w_{n+1}-w^{(0)}
\end{array}\rp=
\lp
\begin{array}{cccc}
R_{11} & 0 & 0 & 0 \\
0 & R_{22} & R_{23} & R_{24} \\
0 & R_{32} & R_{33} & R_{34} \\
0 & R_{42} & R_{43} & R_{44}
\end{array}\rp
\lp
\begin{array}{c}
t_n-t^{(0)} \\
r_n-r^{(0)} \\
u_n-u^{(0)} \\
w_n-w^{(0)}
\end{array}\rp.
\label{ft12}
\ee
We have $E_1=R_{11}=s^{(d+2)/2}$. Other nonzero matrix elements $R_{ij}$
($i=2,3,4$; $j=2,3,4$) and the eigenvalues $E_2$, $E_3$, $E_4$ coincide,
respectively, with the quantities $R_{i_1j_1}$ ($i_1=i-1$; $j_1=j-1$) and
$E_1$, $E_2$, $E_3$ obtained in the case of $h=0$. The quantities $f_0$,
$\varphi_0$, and $\psi_0$ characterizing the fixed-point coordinates
\bea
& & t^{(0)}=0, \qquad r^{(0)}=-f_0\btPhia,\non
& & u^{(0)}=\varphi_0(\btPhia)^2, \qquad
w^{(0)}=\psi_0(\btPhia)^3
\label{ft13}
\eea
as well as the remaining coefficients in Eqs. (\ref{ft10}) are also
defined on the basis of expressions corresponding to a zero
external field.

\section{Using the generalized point of exit of the system from the
critical-regime region for calculating the free energy}

Let us calculate the free energy $F=-kT\ln Z$ of a $3D$ Ising-like
system above the critical temperature $T_c$. The basic idea of such
a calculation on the microscopic level consists in the separate inclusion
of the contributions from short-wave ($F_{CR}$, the region of the
critical regime) and long-wave ($F_{LGR}$, the region of the limiting
Gaussian regime) modes of spin-moment density
oscillations \cite{ymo287,rev9789,ykpmo101}:
\be
F=F_0+F_{CR}+F_{LGR}.
\label{ft14}
\ee
Here $F_0=-kTN\ln 2$ is the free energy of $N$ noninteracting spins.
Each of three components in Eq. (\ref{ft14}) corresponds to
individual factor in the convenient representation
\be
Z=2^N Z_{CR}Z_{LGR}
\label{ft15}
\ee
for the partition function given by Eq. (\ref{ft7}). The
contributions from short- and long-wave modes to the free energy of the
system in the presence of an external field are calculated in the
$\rho^6$ model approximation according to the scheme proposed in
\cite{kpd297,p599,ykp202,ypk102}. Short-wave modes are characterized by
a RG symmetry and are described by the non-Gaussian measure density. The
calculation of the contribution from long-wave modes is based on using
the Gaussian measure density as the basis one. Here, we have developed
a direct method of calculations with the results obtained by taking into
account the short-wave modes as initial parameters. The main results
obtained in the course of deriving the complete expression for the free
energy of the system are presented below.

\subsection{Region of the critical regime}

A calculation technique based on the $\rho^6$ model for the contribution
$F_{CR}$ is similar to that elaborated in the absence of an external
field (see, for example, \cite{ykpmo101,p599,ykp202}). Carrying out the
summation of partial tree energies $F_n$ over the layers of the phase
space of CV, we can calculate $F_{CR}$:
\bea
& & \qquad F_{CR}=F'_0+F'_{CR},\non
& & F'_0=-kTN'[\ln Q(\cM)+\ln Q(d)],\non
& & F'_{CR}=~\sli_{n=1}^{\np}F_n.
\label{ft16}
\eea
An explicit dependence of $F_n$ on the layer number $n$ is obtained using
solutions (\ref{ft10}) of RR and series expansions of special functions
in small deviations of the basic arguments from their values at the fixed
point. The main peculiar feature of the present calculations lies in
using the generalized point of exit of the system from the critical
regime of order-parameter fluctuations. The inclusion of the more
complicated expression for the exit point (as a function of both the
temperature and field variables) \cite{k107}
\be
\np=-\frac{\ln(\hhcqp)}{2\ln E_1}-1
\label{ft17}
\ee
leads to the distinction between formula (\ref{ft16}) for $F'_{CR}$ and
the analogous relation at $h=0$ \cite{p599,ykp202}. The quantity
$\tilde h=h'/f_0$ is determined by the dimensionless field $h'$,
while the quantity $\tilde h_c=\tilde\tau^{p_0}$ is a function
of the reduced temperature $\tau=(T-T_c)/T_c$. Here
$\tilde\tau=\tilde c_1^{(0)}\tau/f_0$, $p_0=\ln E_1/\ln E_2=(d+2)\nu/2$,
$\tilde c_1^{(0)}$ characterizes the coefficient $c_1$ in
solutions (\ref{ft10}) of RR, $\nu=\ln s/ln E_2$ is the critical exponent
of the correlation length. At $h=0$, $\np$ becomes
$\mt=-\ln\tilde\tau/\ln E_2-1$ (see \cite{ykpmo101,p599,ykp202}).
At $T=T_c$ ($\tau=0$), the quantity $\np$ coincides with the exit point
$n_h=-\ln\tilde h/\ln E_1-1$ \cite{kpp204}. The limiting value of the
field $\tilde h_c$ is obtained by the equality of the exit points defined
by the temperature and by the field ($\mt=n_h$).

Having expression (\ref{ft17}) for $\np$, we arrive at
the relations \cite{p907}
\bea
& & E_1^{\np+1}=(\hhcqp)^{-1/2}, \qquad
\tilde\tau E_2^{\np+1}=H_c,\non
& & H_c=\tilde h_c^{1/p_0}(\hhcqp)^{-1/(2p_0)},\non
& & E_3^{\np+1}=H_3, \qquad
H_3=(\hhcqp)^{\Delb/(2p_0)},\non
& & E_4^{\np+1}=H_4, \qquad
H_4=(\hhcqp)^{\Delta_2/(2p_0)},\non
& & s^{-(\np+1)}=(\hhcqp)^{1/(d+2)},
\label{ft18}
\eea
where $\Delb=-\ln E_3/\ln E_2$ and $\Delta_2=-\ln E_4/\ln E_2$ are the
exponents, which determine the first and second confluent corrections,
respectively. Numerical values of the quantities $E_l$ ($l=1,2,3,4$),
$\nu$, $\Delb$, and $\Delta_2$ for the optimal RG parameter $s=s^*=2.7349$
are given in Table~\ref{tab1}.
\begin{table}[htbp]
\caption{The eigenvalues $E_l$ and the exponents $\nu$,
$\Delb$, $\Delta_2$ for the $\rho^6$ model.}
\label{tab1}
\begin{center}
\begin{tabular}{lllllll}
\hline
\multicolumn{1}{c}{$E_1$} & \multicolumn{1}{c}{$E_2$} &
\multicolumn{1}{c}{$E_3$} & \multicolumn{1}{c}{$E_4$} &
\multicolumn{1}{c}{$\nu$} & \multicolumn{1}{c}{$\Delb$} &
\multicolumn{1}{c}{$\Delta_2$} \\
\hline
12.3695 & 4.8468 & 0.4367 & 0.0032 & 0.637 & 0.525 & 3.647 \\
\hline
\end{tabular}
\end{center}
\end{table}
In the weak-field
region ($\tilde h\ll \tilde h_c$), quantities (\ref{ft18}) can be
calculated with the help of the following expansions:
\bea
& & E_1^{\np+1}=\tilde h_c^{-1}\lp 1-\half~\hhcqf\rp, \qquad
\tilde h_c^{-1}=\tilde\tau^{-p_0},\non
& & H_c=1-\frac{1}{2p_0}~\hhcqf,\non
& & H_3=\tilde h_c^{\Delb/p_0}\lp 1+\frac{\Delb}{2p_0}~\hhcqf\rp, \qquad
\tilde h_c^{\Delb/p_0}=\tilde\tau^{\Delb},\non
& & H_4=\tilde h_c^{\Delta_2/p_0}\lp 1+\frac{\Delta_2}{2p_0}~\hhcqf\rp,
\qquad \tilde h_c^{\Delta_2/p_0}=\tilde\tau^{\Delta_2},\non
& & s^{-(\np+1)}=\tilde h_c^{2/(d+2)}\lp 1+\frac{1}{d+2}~\hhcqf\rp,\non
& & \tilde h_c^{2/(d+2)}=\tilde\tau^{\nu}.
\label{ft19}
\eea
In the strong-field region ($\tilde h\gg \tilde h_c$), these quantities
satisfy the expressions
\bea
& & E_1^{\np+1}=\tilde h^{-1}\lp 1-\half~\hchqf\rp,\non
& & H_c=(\tilde h_c/\tilde h)^{1/p_0}\lp 1-\frac{1}{2p_0}~\hchqf\rp,\non
& & H_3=\tilde h^{\Delb/p_0}\lp 1+\frac{\Delb}{2p_0}~\hchqf\rp,\non
& & H_4=\tilde h^{\Delta_2/p_0}\lp 1+\frac{\Delta_2}{2p_0}~\hchqf\rp,\non
& & s^{-(\np+1)}=\tilde h^{2/(d+2)}\lp 1+\frac{1}{d+2}~\hchqf\rp.
\label{ft20}
\eea
It should be noted that the variables $\tilde h/\tilde h_c$ (the weak
fields) and $(\tilde h_c/\tilde h)^{1/p_0}$ (the strong fields) coincide
with the accepted choice of the arguments for scaling functions
in accordance with the scaling theory. In the particular case of $h=0$
and $\tau\neq 0$, Eqs. (\ref{ft19}) are defined as
$E_1^{\np+1}=\tilde\tau^{-p_0}$, $H_c=1$, $H_3=\tilde\tau^{\Delb}$,
$H_4=\tilde\tau^{\Delta_2}$, $s^{-(\np+1)}=\tilde\tau^{\nu}$. At
$h\neq 0$ and $\tau=0$, we have $\tilde h E_1^{\np+1}=1$, $H_c=0$,
$H_3=\tilde h^{\Delb/p_0}$, $H_4=\tilde h^{\Delta_2/p_0}$,
$s^{-(\np+1)}=\tilde h^{2/(d+2)}$ [see Eqs. (\ref{ft20})].

We shall perform the further calculations on the basis of Eqs.
(\ref{ft18}), which are valid in the general case for the regions of
small, intermediate (the crossover region), and large field values. The
inclusion of $E_3^{\np+1}$ (or $H_3$) leads to the formation of the first
confluent corrections in the expressions for thermodynamic characteristics
of the system. The quantity $E_4^{\np+1}$ (or $H_4$) is responsible for
the emergence of the second confluent corrections. The cases of the weak
or strong fields can be obtained from general expressions by using
Eqs. (\ref{ft19}) or (\ref{ft20}). We disregard the second confluent
correction in our calculations. This is due to the fact that the
contribution from the first confluent correction to thermodynamic
functions near the critical point ($\tau=0$, $h=0$) for various values
of $s$ is more significant than the small contribution from the second
correction ($\hhcqp\ll 1$, $\Delb$ is of the order of 0.5, and
$\Delta_2 > 2$, see Table~\ref{tab1} and \cite{kpd297}).

Proceeding from an explicit dependence of $F_n$ on the layer number $n$
\cite{ykpmo101,kpd297,p599} and taking into account Eqs.
(\ref{ft18}), we can now write the final expression for
$F_{CR}$ (\ref{ft16}):
\bea
& & F_{CR}=-kTN'\lp \gamma_0^{(CR)}+\gamma_1\tau+\gamma_2\tau^2\rp+
F_s,\non
& & F_s=kTN's^{-3(\np+1)}\lp \bar \gamma_3^{(CR)(0)+}+
\bar \gamma_3^{(CR)(1)+}\ccaaHd\rp.\non
\label{ft21}
\eea
Here $c_{20}^{(0)}$ characterizes $c_2$ in solutions (\ref{ft10}) of RR,
\bea
\bar \gamma_3^{(CR)(0)+}&=&\frac{f_{CR}^{(0)}}{1-s^{-3}}+
\frac{f_{CR}^{(1)}\varphi_0^{-1/2}\faHc}{1-E_2 s^{-3}}\non
& & +\frac{f_{CR}^{(7)}\varphi_0^{-1}(\faHc)^2}{1-E_2^2 s^{-3}},\non
\bar \gamma_3^{(CR)(1)+}&=&\frac{f_{CR}^{(2)}\varphi_0^{-1}}
{1-E_3 s^{-3}}+\frac{f_{CR}^{(4)}\varphi_0^{-3/2}\faHc}
{1-E_2E_3 s^{-3}}\non
& & +\frac{f_{CR}^{(8)}\varphi_0^{-2}(\faHc)^2}{1-E_2^2E_3 s^{-3}},
\label{ft22}
\eea
and the coefficients
\bea
& & \gamma_0^{(CR)}=\gamma_0^{(0)}+\delta_0^{(0)},\non
& & \gamma_k=\gamma_0^{(k)}+\delta_0^{(k)}, \qquad k=1,2
\label{ft23}
\eea
are determined by the components of the quantities
\bea
& & \gamma_0=\gamma_0^{(0)}+\gamma_0^{(1)}\tau+\gamma_0^{(2)}\tau^2,\non
& & \delta_0=\delta_0^{(0)}+\delta_0^{(1)}\tau+\delta_0^{(2)}\tau^2.
\label{ft24}
\eea
The components $\delta_0^{(i)}$ ($i=0,1,2$) satisfy the earlier relations
\cite{ykpmo101,kpd297,p599} obtained in the case of a zero external
field. The components $\gamma_0^{(i)}$ are given by the corresponding
expressions at $h=0$ under condition that the eigenvalues $E_1$, $E_2$,
and $E_3$ should be replaced by $E_2$, $E_3$, and $E_4$, respectively.

Let us now calculate the contribution to the free energy of the system
from the layers of the CV phase space beyond the point of exit from the
critical-regime region. The calculations are performed according to the
scheme proposed in \cite{ymo287,ykpmo101,ykp202,ypk102}. As in the
previous study, while calculating the partition function component
$Z_{LGR}$ from Eq. (\ref{ft15}), it is convenient to single out two
regions of values of wave vectors. The first is the transition region
($Z_{LGR}^{(1)}$) corresponding to values of $\ve{k}$ close to $B_{\np}$,
while the second is the Gaussian region ($Z_{LGR}^{(2)}$) corresponding
to small values of wave vector ($k\rightarrow 0$). Thus, we have
\be
Z_{LGR}=Z_{LGR}^{(1)}Z_{LGR}^{(2)}.
\label{ft25}
\ee

\subsection{Transition region}

This region corresponds to $\tilde m_0$ layers of the phase space of CV.
The lower boundary of the transition region is determined by the point of
exit of the system from the critical-regime region ($n=\np+1$). The upper
boundary corresponds to the layer $\np+\tilde m_0+1$. We use for
$\tilde m_0$ the integer closest to $\tilde m'_0$. The condition for
obtaining $\tilde m'_0$ is the equality \cite{p599,ykp202}
\be
\mid h_{\np+\tilde m'_0}\mid=\frac{A_0}{1-s^{-3}},
\label{ft26}
\ee
where $A_0$ is a large number ($A_0\geq 10$).

The free energy contribution
\bea
F_{LGR}^{(1)}&=&-kTN_{\np+1}~\sli_{m=0}^{\tilde m_0}s^{-3m}
f_{LGR_1}(m),\non
f_{LGR_1}(m)&=&\ln\lp\frac{2}{\pi}\rp+\frac{1}{4}~\ln 24-
\frac{1}{4}~\ln C(\eta_{\npm},\xi_{\npm})\non
& & +\ln I_0(h_{\npmb},\alpha_{\npmb})\non
& & +\ln I_0(\eta_{\npm},\xi_{\npm})
\label{ft27}
\eea
corresponding to $Z_{LGR}^{(1)}$ from Eq. (\ref{ft25}) is calculated
by using the solutions of RR.

\begin{figure}[htbp]
\centering \includegraphics[width=0.4953\textwidth]{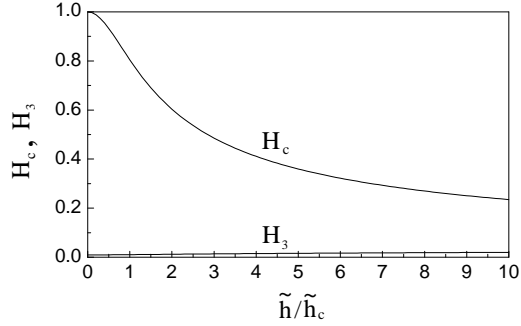}
\caption{Dependence of quantities $H_c$ and $H_3$ on
the ratio $\tilde h/\tilde h_c$ for the RG parameter
$s=s^*=2.7349$ and the reduced temperature $\tau=10^{-4}$.}
\label{fig1}
\end{figure}

The basic arguments in the ($\npm$)th layer
\bea
& & h_{\npm}=(r_{\npm}+q)(6/u_{\npm})^{1/2},\non
& & \alpha_{\npm}=\frac{\sqrt 6}{15}~w_{\npm}/u_{\npm}^{3/2}
\label{ft28}
\eea
can be presented using the relations
\bea
t_{\npm}&=&-s_0^{d/2}f_0E_1^{m-1}\tilde h (\hhcqp)^{-1/2},\non
r_{\npm}&=&\btPhia\Bigl( -f_0+\faHc E_2^{m-1}\non
& & +\ccaaHd \varphi_0^{-1/2}w_{12}^{(0)}E_3^{m-1}\Bigr),\non
u_{\npm}&=&(\btPhia)^2\Bigl( \varphi_0+
\faHc \varphi_0^{1/2}w_{21}^{(0)}E_2^{m-1}\non
& & +\ccaaHd E_3^{m-1}\Bigr),\non
w_{\npm}&=&(\btPhia)^3\Bigl( \psi_0+\faHc \varphi_0w_{31}^{(0)}
E_2^{m-1}\non
& & +\ccaaHd \varphi_0^{1/2}w_{32}^{(0)}E_3^{m-1}\Bigr)
\label{ft29}
\eea
obtained on the basis of Eqs. (\ref{ft10}) and (\ref{ft18}). We
arrive at the following expressions:
\bea
& & h_{\npm}=h_{\npm}^{(0)}\lp 1+
\bar h_{\npm}^{(1)}\ccaaHd\rp,\non
h_{\npm}^{(0)}&=&\sqrt 6~\frac{\bar q-f_0+\faHc E_2^{m-1}}
{(\varphi_0+\faHc \varphi_0^{1/2}w_{21}^{(0)}E_2^{m-1})^{1/2}},\non
\bar h_{\npm}^{(1)}&=&E_3^{m-1}\Biggl(
\frac{\varphi_0^{-1/2}w_{12}^{(0)}}
{\bar q-f_0+\faHc E_2^{m-1}}\non
& & -\half~\frac{1}
{\varphi_0+\faHc \varphi_0^{1/2}w_{21}^{(0)}E_2^{m-1}}\Biggr);\non
& & \alpha_{\npm}=\alpha_{\npm}^{(0)}\lp 1+
\bar \alpha_{\npm}^{(1)}\ccaaHd\rp,\non
\alpha_{\npm}^{(0)}&=&\frac{\sqrt 6}{15}~
\frac{\psi_0+\faHc \varphi_0w_{31}^{(0)}E_2^{m-1}}
{(\varphi_0+\faHc \varphi_0^{1/2}w_{21}^{(0)}E_2^{m-1})^{3/2}},\non
\bar \alpha_{\npm}^{(1)}&=&E_3^{m-1}\Biggl(
\frac{\varphi_0^{1/2}w_{32}^{(0)}}
{\psi_0+\faHc \varphi_0w_{31}^{(0)}E_2^{m-1}}\non
& & -\frac{3}{2}~\frac{1}
{\varphi_0+\faHc \varphi_0^{1/2}w_{21}^{(0)}E_2^{m-1}}\Biggr).
\label{ft30}
\eea
In contrast to $H_c$, the quantity $H_3$ in expressions (\ref{ft30}) for
$h_{\npm}$ and $\alpha_{\npm}$ as well as in expression (\ref{ft21}) for
$F_s$ takes on small values with the variation of the field $\tilde h$
(see Fig.~\ref{fig1}).
The quantity $H_c$ at $\tilde h\rightarrow 0$
and near $\tilde h_c$ is close to unity and series expansions in $H_c$
are not effective here.

Power series in small deviations $(h_{\npm}-h_{\npm}^{(0)})$ and
$(\alpha_{\npm}-\alpha_{\npm}^{(0)})$ for the special functions appearing
in the expressions for the intermediate arguments
\bea
\eta_{\npm}&=&(6s^d)^{1/2}F_2(h_{\npm},\alpha_{\npm})\non
& & \times\lbr C(h_{\npm},\alpha_{\npm})\rbr^{-1/2},\non
\xi_{\npm}&=&\frac{\sqrt 6}{15}~s^{-d/2}N(h_{\npm},\alpha_{\npm})\non
& & \times\lbr C(h_{\npm},\alpha_{\npm})\rbr^{-3/2}
\label{ft31}
\eea
allow us to find the relations
\bea
\eta_{\npm}&=&\eta_{\npm}^{(0)}\Bigl[ 1-\Bigl( \bar \eta_1^{(\npm)}
h_{\npm}^{(0)}\bar h_{\npm}^{(1)}\non
& & +\bar \eta_2^{(\npm)}\alpha_{\npm}^{(0)}
\bar \alpha_{\npm}^{(1)}\Bigr)\ccaaHd\Bigr],\non
\xi_{\npm}&=&\xi_{\npm}^{(0)}\Bigl[ 1-\Bigl( \bar \xi_1^{(\npm)}
h_{\npm}^{(0)}\bar h_{\npm}^{(1)}\non
& & +\bar \xi_2^{(\npm)}\alpha_{\npm}^{(0)}
\bar \alpha_{\npm}^{(1)}\Bigr)\ccaaHd\Bigr].
\label{ft32}
\eea
The quantities $\eta_{\npm}^{(0)}$, $\bar \eta_1^{(\npm)}$,
$\bar \eta_2^{(\npm)}$ and $\xi_{\npm}^{(0)}$, $\bar \xi_1^{(\npm)}$,
$\bar \xi_2^{(\npm)}$ are functions of
$F_{2l}^{*(\npm)}=I_{2l}^{*(\npm)}/I_0^{*(\npm)}$, where
\be
I_{2l}^{*(\npm)}=\ili_0^{\infty}x^{2l}\exp(-h_{\npm}^{(0)}x^2-
x^4-\alpha_{\npm}^{(0)}x^6)~dx.
\label{ft33}
\ee

Proceeding from expression (\ref{ft27}) for $f_{LGR_1}(m)$, we can now
write the following relation accurate to within $H_3$:
\bea
& &  f_{LGR_1}(m)=f_{LGR_1}^{(0)}(m)+\bar f_{LGR_1}^{(1)}(m)
\ccaaHd,\non
f_{LGR_1}^{(0)}(m)&=&\ln\lp\frac{2}{\pi}\rp+\frac{1}{4}~\ln 24-
\frac{1}{4}~\ln C(\eta_{\npm}^{(0)},\xi_{\npm}^{(0)})\non
& & +\ln I_0(h_{\npmb}^{(0)},\alpha_{\npmb}^{(0)})\non
& & +\ln I_0(\eta_{\npm}^{(0)},\xi_{\npm}^{(0)}),\non
\bar f_{LGR_1}^{(1)}(m)&=&\varphi_1^{(\npm)}h_{\npm}^{(0)}
\bar h_{\npm}^{(1)}\non
& & +\varphi_2^{(\npm)}\alpha_{\npm}^{(0)}
\bar \alpha_{\npm}^{(1)}\non
& & +\varphi_3^{(\npmb)}h_{\npmb}^{(0)}\bar h_{\npmb}^{(1)}\non
& & +\varphi_4^{(\npmb)}\alpha_{\npmb}^{(0)}\bar \alpha_{\npmb}^{(1)},\non
\varphi_k^{(\npm)}&=&b_k^{(\npm)}+P_{4k}^{(\npm)}/4, \qquad k=1,2,\non
\varphi_3^{(\npmb)}&=&-F_2^{*(\npmb)},\non
\varphi_4^{(\npmb)}&=&-F_6^{*(\npmb)}.
\label{ft34}
\eea
The quantities $b_k^{(\npm)}$, $P_{4k}^{(\npm)}$ depend on
$F_{2l}^{*(\npm)}$ as well as on
$F_{2l}^{**(\npm)}=I_{2l}^{**(\npm)}/I_0^{**(\npm)}$, where
\be
I_{2l}^{**(\npm)}=\ili_0^{\infty}x^{2l}\exp(-\eta_{\npm}^{(0)}x^2-
x^4-\xi_{\npm}^{(0)}x^6)~dx.
\label{ft35}
\ee

The final result for $F_{LGR}^{(1)}$ [see Eqs. (\ref{ft27}) and
(\ref{ft34})] assumes the form
\bea
& & F_{LGR}^{(1)}=-kTN's^{-3(\np+1)}\lp \bar f_{TR}^{(0)}+
\bar f_{TR}^{(1)}\ccaaHd\rp,\non
& & \qquad \bar f_{TR}^{(0)}=\sli_{m=0}^{\tilde m_0}
s^{-3m}f_{LGR_1}^{(0)}(m),\non
& & \qquad \bar f_{TR}^{(1)}=\sli_{m=0}^{\tilde m_0}
s^{-3m}\bar f_{LGR_1}^{(1)}(m).
\label{ft36}
\eea
On the basis of Eqs. (\ref{ft26}) and (\ref{ft30}), it is possible
to obtain the quantity $\tilde m'_0$ determining the summation limit
$\tilde m_0$ in formulas (\ref{ft36}):
\bea
& & \tilde m'_0=\frac{\ln L_0-\ln H_c}{\ln E_2}+1,\non
& & L_0=A_1+(A_1^2-A_2)^{1/2},\non
& & A_1=1-\frac{\bar q}{f_0}+\frac{A_0^2\varphi_0^{1/2}w_{21}^{(0)}}
{12f_0(1-s^{-3})^2},\non
& & A_2=1-2\frac{\bar q}{f_0}+\lp\frac{\bar q}{f_0}\rp^2-
\frac{A_0^2\varphi_0}{6f_0^2(1-s^{-3})^2}.
\label{ft37}
\eea

Let us now calculate the contribution to the free energy of the system
from long-wave modes in the range of wave vectors
\bea
& & k\leq B's^{-\nps},\non
& & \nps=\np+\tilde m_0+2
\label{ft38}
\eea
using the Gaussian measure density.

\subsection{Region of small values of wave vector ($k\rightarrow 0$)}

The free energy component
\be
F_{LGR}^{(2)}=\half kT\lbr N_{\nps}\ln P_2^{(\npsb)}+\sli_{k=0}^{B_{\nps}}
\ln \tilde d_{\nps}(k)-\frac{N(h')^2}{\tilde d_{\nps}(0)}\rbr
\label{ft39}
\ee
corresponding to $Z_{LGR}^{(2)}$ from Eq. (\ref{ft25}) is similar
to that presented in \cite{ykpmo101,p599,ykp202}. The calculations of
the first and second terms in Eq. (\ref{ft39}) are associated with
the calculations of the quantities
\bea
\!\!\!P_2^{(\npsb)}&=&2h_{\npsb}F_2(h_{\npsb},\alpha_{\npsb})\non
& & \times\lbr d_{\npsb}(B_{\nps},B_{\npsb})\rbr^{-1},\non
\tilde d_{\nps}(k)&=&\lbr P_2^{(\npsb)}\rbr^{-1}\!+\!\beta
\tilde{\Phi}(B_{\nps},B_{\npsb})\!-\!\btPhik,
\label{ft40}
\eea
where
\be
d_{\npsb}(B_{\nps},B_{\npsb})=s^{-2(\npsb)}(r_{\npsb}+q),
\label{ft41}
\ee
and $r_{\npsb}$, $h_{\npsb}=h_{\npsb}^{(0)}\lp 1+
\bar h_{\npsb}^{(1)}\ccaaHd\rp$, $\alpha_{\npsb}=\alpha_{\npsb}^{(0)}
\lp 1+\bar \alpha_{\npsb}^{(1)}\ccaaHd\rp$ satisfy the corresponding
expressions from Eqs. (\ref{ft29}) and (\ref{ft30}) at
$m=\tilde m_0+1$.

Introducing the designation
\be
p=h_{\npsb}F_2(h_{\npsb},\alpha_{\npsb})
\label{ft42}
\ee
and presenting it in the form
\be
p^{-1}=p_0(1+\bar p_1\ccaaHd),
\label{ft43}
\ee
we obtain the following relations for the coefficients:
\bea
p_0&=&\lbr h_{\npsb}^{(0)}p_{20}^{(\npsb)}\rbr^{-1},\non
\bar p_1&=&-\bar h_{\npsb}^{(1)}\lp 1-
p_{21}^{(\npsb)}h_{\npsb}^{(0)}\rp\non
& & +p_{22}^{(\npsb)}\alpha_{\npsb}^{(0)}\bar \alpha_{\npsb}^{(1)}.
\label{ft44}
\eea
The quantities
\bea
& & p_{20}^{(\npsb)}=F_2^{*(\npsb)}, \quad
p_{21}^{(\npsb)}=\frac{F_4^{*(\npsb)}}{F_2^{*(\npsb)}}-
F_2^{*(\npsb)},\non
& & p_{22}^{(\npsb)}=\frac{F_8^{*(\npsb)}}{F_2^{*(\npsb)}}-
F_6^{*(\npsb)}
\label{ft45}
\eea
determine the function
\bea
F_2(h_{\npsb},\alpha_{\npsb})&=&p_{20}^{(\npsb)}\lbr 1-
\lp p_{21}^{(\npsb)}h_{\npsb}^{(0)}\bar h_{\npsb}^{(1)}\rd\rd\non
& & \ld\ld\!\!+p_{22}^{(\npsb)}\alpha_{\npsb}^{(0)}
\bar \alpha_{\npsb}^{(1)}\rp\ccaaHd\rbr.
\label{ft46}
\eea
Here $F_{2l}^{*(\npsb)}=I_{2l}^{*(\npsb)}/I_0^{*(\npsb)}$, where
\be
I_{2l}^{*(\npsb)}=\ili_0^{\infty}x^{2l}\exp(-h_{\npsb}^{(0)}x^2-
x^4-\alpha_{\npsb}^{(0)}x^6)~dx.
\label{ft47}
\ee

Taking into account Eqs. (\ref{ft41}) and (\ref{ft43}), we rewrite
formulas (\ref{ft40}) as
\bea
P_2^{(\npsb)}&=&\Biggl\{ \half~s^{-2(\npsb)}\btPhia p_0
(\bar q-f_0+\faHc E_2^{\tilde m_0})\non
& & \times\Biggl[ 1\!+\!\lp\frac{\varphi_0^{-1/2}w_{12}^{(0)}E_3^{\tilde m_0}}
{\bar q-f_0+\faHc E_2^{\tilde m_0}}\!+\!
\bar p_1\rp\ccaaHd\Biggr]\Biggr\}^{-1},\non
\tilde d_{\nps}(k)&=&s^{-2(\npsb)}\btPhia\tilde G+
2\btPhia b^2 k^2,\non
\tilde G&=&g_0(1+\bar g_1\ccaaHd),\non
g_0&=&\half~\lbr (-f_0+\faHc E_2^{\tilde m_0})p_0+
(p_0-2)\bar q\rbr,\non
\bar g_1&=&\half~\frac{p_0}{g_0}\bigr[ \bar p_1
(\bar q-f_0+\faHc E_2^{\tilde m_0})\non
& & +\varphi_0^{-1/2}w_{12}^{(0)}E_3^{\tilde m_0}\bigr].
\label{ft48}
\eea
The second term in Eq. (\ref{ft39}) is defined by the expression
\bea
\half~\sli_{k=0}^{B_{\nps}}\ln \tilde d_{\nps}(k)&=&N_{\nps}
\Bigl\{ \half~\ln(\tilde G+s^{-2})+\ln s-\nps\ln s\non
& & +\half~\ln (\btPhia)-\frac{1}{3}
+\tilde G s^2\non
& & -(\tilde G s^2)^{3/2}
\arctan \lbr (\tilde G s^2)^{-1/2}\rbr\Bigr\}.
\label{ft49}
\eea

Relations (\ref{ft48}) and (\ref{ft49}) make it possible to find
the component $F_{LGR}^{(2)}$ in the form
\bea
F_{LGR}^{(2)}&=&-kT\biggl[ N's^{-3(\np+1)}(\bar f^{(0)'}+
\bar f^{(1)'}\ccaaHd)\non
& & +\frac{N(h')^2\bar \gamma_4^+}{\btPhia}~s^{2(\np+1)}
(1-\bar g_1\ccaaHd)\biggr],\non
& & \bar f^{(0)'}=s^{-3(\tilde m_0+1)}f^{(0)}, \quad
\bar f^{(1)'}=s^{-3(\tilde m_0+1)}\bar f^{(1)},\non
f^{(0)}&=&-\half~\ln\lp \frac{s^{-2}+g_0}{g_0+\bar q}\rp+\frac{1}{3}\non
& & -g'_0\lbr 1-\sqrt{g'_0}~\arctan\lp \frac{1}{\sqrt{g'_0}}\rp\rbr,\non
\bar f^{(1)}&=&\half~\lp \frac{g_0\bar g_1}{g_0+\bar q}-
\frac{\bar g_1}{(g'_0)^{-1}+1}-\frac{g'_0\bar g_1}{(g'_0)^{-1}+1}\rp\non
& & -g'_0\bar g_1\lbr 1-\frac{3}{2}~\sqrt{g'_0}~\arctan\lp \frac{1}
{\sqrt {g'_0}}\rp\rbr,\non
g'_0&=&s^2g_0, \qquad \bar \gamma_4^+=s^{2\tilde m_0}/(2g_0).
\label{ft50}
\eea

On the basis of Eqs. (\ref{ft36}) and (\ref{ft50}), we can write the
following expression for the general contribution
$F_{LGR}=F_{LGR}^{(1)}+F_{LGR}^{(2)}$ to the free energy of the system
from long-wave modes of spin-moment density oscillations:
\bea
F_{LGR}&=&-kT\biggl[ N's^{-3(\np+1)}(\bar f_{LGR}^{(0)}+
\bar f_{LGR}^{(1)}\ccaaHd)\non
& & +\frac{N(h')^2\bar \gamma_4^+}
{\btPhia}~s^{2(\np+1)}(1-\bar g_1\ccaaHd)\biggr],\non
& & \bar f_{LGR}^{(l)}=\bar f_{TR}^{(l)}+\bar f^{(l)'}, \qquad
l=0,1.
\label{ft51}
\eea

\section{Total free energy of the system at $T>T_c$ as function of
temperature, field and microscopic parameters}

The total free energy of the system is calculated taking into account
Eqs. (\ref{ft14}), (\ref{ft21}), and (\ref{ft51}). Collecting the
contributions to the free energy from all regimes of fluctuations at
$T>T_c$ in the presence of an external field and using the relation
for $s^{-(\np+1)}$ from Eqs. (\ref{ft18}), we obtain
\bea
F&\!=\!&-kTN\Biggl[ \gamma'_0+\gamma'_1\tau+\gamma'_2\tau^2+
(\bar \gamma_3^{(0)+}+\bar \gamma_3^{(1)+}\ccaaHd)\non
& & \times\!(\hhcqp)^{3/5}
\!+\!\frac{\bar \gamma_4^+(h')^2}{\btPhia}(1\!-\!\bar g_1\ccaaHd)
(\hhcqp)^{-2/5}\Biggr],\non
& & \gamma'_0=\ln 2+s_0^{-3}\gamma_0^{(CR)}, \quad
\gamma'_1=s_0^{-3}\gamma_1, \quad \gamma'_2=s_0^{-3}\gamma_2,\non
& & \bar \gamma_3^{(l)+}=s_0^{-3}(-\bar \gamma_3^{(CR)(l)+}+
\bar f_{LGR}^{(l)}), \qquad l=0,1.
\label{ft52}
\eea
The coefficients $\gamma_0^{(CR)}$, $\gamma_1$, $\gamma_2$ are defined
by Eqs. (\ref{ft23}), $\bar g_1$ is presented in Eqs.
(\ref{ft48}), and $\bar \gamma_4^+$ is given in Eqs. (\ref{ft50}).
The coefficients of the non-analytic component of the free energy $F$
[see Eqs. (\ref{ft52})] depend on $H_c$. The terms proportional to
$H_3$ determine the confluent corrections by the temperature and field.
As is seen from the expression for $F$, the free energy of the system at
$\tilde h=0$ and $\tilde\tau=0$, in addition to terms proportional to
$\tilde\tau^{3\nu}$ (or $\tilde h_c^{6/5}$) and $\tilde h^{6/5}$,
contains the terms proportional to $\tilde\tau^{3\nu+\Delb}$ and
$\tilde h^{6/5+\Delb/p_0}$, respectively. At $\tilde h\neq 0$ and
$\tilde\tau\neq 0$, the terms of both types are present. It should be
noted that $\Delb>\Delb/p_0$. At $\tilde h=\tilde h_c$,
we have $\tilde\tau^{\Delb}=\tilde h^{\Delb/p_0}$ and the contributions
to the thermodynamic characteristics of the system from both types of
the corrections become of the same order.

The advantage of the method presented in this article is the possibility
of deriving analytic expressions for the free-energy coefficients as
functions of the microscopic parameters of the system (the lattice
constant $c$ and parameters of the interaction potential, i.e., the
effective radius $b$ of the potential, the Fourier transform $\tPhia$
of the potential for $k=0$).

\section{Conclusions}

An analytic method for calculating the total free energy of a $3D$
Ising-like system (a $3D$ uniaxial magnet) near the critical point is
developed on the microscopic level in the higher non-Gaussian
approximation based on the sextic distribution for modes of spin-moment
density oscillations (the $\rho^6$ model). The simultaneous effect of
the temperature and field on the behavior of the system is taken into
account. An external field is introduced in the Hamiltonian of the
system from the outset. In contrast to previous studies on the basis of
the asymmetric $\rho^4$ model \cite{pkp105,kpp406,p806}, the field in
the initial process of calculating the partition function of the system
is not included in the Jacobian of transition from the set of spin
variables to the set of CV. Such an approach leads to the appearance of
the first, second, fourth, and sixth powers of CV in the expression for
the partition function and allows us to simplify the mathematical
description because the odd part is represented only by the linear term.

The theory is being built ab initio beginning from
the Hamiltonian of the system up to the expression for
the free energy. The main distinctive feature of the proposed method is
the separate inclusion of the contributions to the free energy from the
short- and long-wave spin-density oscillation modes. The generalized
point of exit of the system from the critical regime contains both the
temperature and field variables. The form of the temperature and field
dependences for the free energy of the system is determined by solutions
of RR near the fixed point. The expression for the free energy $F$
obtained at temperatures $T>T_c$ without using power series in the
scaling variable and without any adjustable parameters can be employed
in the field region near $\tilde h_c$ (the crossover region).
The limiting field $\tilde h_c$ satisfies the condition of the equality
of sizes of the critical-regime region by the temperature and field (the
effect of the temperature and field on the system in the vicinity of the
critical point is equivalent) \cite{pkp105,kpp406,kpp204,p806}. In the
vicinity of $\tilde h_c$, the scaling variable is of the order of unity
and power series in this variable are not effective. We hope that
the proposed method as well as our explicit representations may provide
useful benchmarks in studying the effect of an external magnetic field on
the critical behavior of $3D$ Ising-like systems within the framework of
the higher non-Gaussian approximation (the $\rho^6$ model). Proceeding
from the expression for the free energy, which involves the leading terms
and terms determining the temperature and field confluent corrections, we
can find other thermodynamic characteristics (the average spin moment,
susceptibility, entropy, and specific heat) by direct differentiation of
$F$ with respect to field or temperature.

\end{document}